\documentclass[preprint2]{aastex}

\usepackage{graphics}
\usepackage{amssymb}
\usepackage{aas_macros}

\shorttitle{Migration of a moonlet in a ring of solid particles}
\shortauthors{Crida et al.}

\newcommand{\Sect}[1]{Sect.~\ref{#1}}
\newcommand{\Fig}[1]{Fig.~\ref{#1}}

\newcommand{\Eq}[1]{Eq.~(\ref{#1})}

\begin{document}




\title{Migration of a moonlet in a ring of solid particles\,:\\
       Theory and application to Saturn's propellers.}


\author{Aur\'elien \textsc{Crida}}
\affil{Department of Applied Mathematics and Theoretical Physics,
  University of Cambridge,\\Centre for Mathematical Sciences,
  Wilberforce Road, Cambridge CB3 0WA, UK}
\affil{Laboratoire Cassiop\'ee, Universit\'e de Nice
  Sophia-antipolis / CNRS / Observatoire de la C\^ote d'Azur,\\ 
  B.P. 4229, 06304 Nice Cedex 4, \textsc{France}\ \ \ {\tt crida@oca.eu}}
\author{John~C.~B. \textsc{Papaloizou}}
\affil{Department of Applied Mathematics and Theoretical Physics,
  University of Cambridge,\\Centre for Mathematical Sciences,
  Wilberforce Road, Cambridge CB3 0WA, UK}
\author{Hanno \textsc{Rein}}
\affil{Department of Applied Mathematics and Theoretical Physics,
  University of Cambridge,\\Centre for Mathematical Sciences,
  Wilberforce Road, Cambridge CB3 0WA, UK}
\author{S\'ebastien \textsc{Charnoz}}
\affil{Laboratoire AIM-UMR 7158, CEA/CNRS/Universit\'e Paris Diderot,
     IRFU/Service d'Astrophysique, CEA/Saclay,
     91191 Gif-sur-Yvette Cedex, \textsc{France}}
\author{Julien \textsc{Salmon}}
\affil{Laboratoire AIM-UMR 7158, CEA/CNRS/Universit\'e Paris Diderot,
     IRFU/Service d'Astrophysique, CEA/Saclay,
     91191 Gif-sur-Yvette Cedex, \textsc{France}}

\begin{abstract}

Hundred meter sized objects have been identified by the Cassini
spacecraft in Saturn's A~ring through the so-called ``propeller''
features they create in the ring. These moonlets should migrate, due
to their gravitational interaction with the ring\,; in fact, some
orbital variation have been detected. The standard theory of type~I
migration of planets in protoplanetary disks can't be applied to the
ring system, as it is pressureless. Thus, we compute the differential
torque felt by a moonlet embedded in a two-dimensional disk of solid
particles, with flat surface density profile, both analytically and
numerically. We find that the corresponding migration rate is too
small to explain the observed variations of the propeller's orbit in
Saturn's A-ring.

However, local density fluctuations (due to gravity wakes in the
marginally gravitationally stable A-ring) may exert a stochastic
torque on a moonlet. Our simulations show that this torque can be
large enough to account for the observations, depending on the
parameters of the rings. We find that on time scales of several years
the migration of propellers is likely to be dominated by stochastic
effects (while the former, non-stochastic migration dominates after
$\sim 10^{4-5}$ years). In that case, the migration rates provided by
observations so far suggests that the surface density of the A~ring
should be of the order of $700$~kg\,m$^{-2}$. The age of the
propellers shouldn't exceed $1$ to $100$ million years, depending
on the dominant migration regime.
\end{abstract}




\section{Introduction}
\label{sec:intro}

The theory of disk-satellite interactions has for the most part been
developed after Voyager's encounter with Saturn. The satellites that
orbit beyond the outer edge of the rings perturb the dynamics of the
particles composing the rings. This leads to an exchange of angular
momentum between the rings and the satellites, and to the formation of
density waves in the rings. \citet{LinPapaloizou1979} and
\citet{GT79,GT80} have calculated, using two different methods, the
total torque exerted by a satellite on a disk interior (or exterior)
to its orbit. This torque is called the \emph{one-sided Lindblad
  torque}, because it can be computed as the sum of the torques
exerted at Lindblad resonances with the secondary body. In the lowest
order, local approximation, the inner and outer torques are equal and
opposite\,: the torque exerted by a satellite on a disk located inside
its orbit is negative, with the same absolute value as the positive
torque exerted on a disk located outside the orbit.

Reciprocally, a disk exerts a torque on the secondary body. When the
strictly local approximation is relaxed, the inner and outer torques
are not exactly equal and opposite \citep{Ward1986}. Their sum, called
the \emph{differential Lindblad torque}, is generally negative. As a
consequence, the orbital angular momentum of a body embedded in a
disk decreases, and so does its semi-major axis (on circular
Keplerian orbits, the orbital angular momentum is proportional to the
square root of the semi-major axis). This is planetary migration of
type~I \citep{Ward1997}. So far, this phenomenon has been mainly
studied in the frame of planets embedded in protoplanetary gaseous
disks \citep[see][for a review]{Papaloizou-etal-2007}.

The Cassini spacecraft has been orbiting the Saturnian ring system
since 2004, offering the possibility to observe the coupled evolution
of the ring system and the satellites. Due to short orbital timescales
(1 year is equivalent to about 700 orbits of the A~ring) it may be
possible to observe the exchange of angular momentum between the two
systems. One of the most striking discoveries of the Cassini
spacecraft is the observation of propeller shaped features in the
A~ring (located between $122\,000$ and $137\,000$~km from Saturn),
with longitudinal extent about $3$~km
\citep{Tiscareno-etal-2006,Tiscareno-etal-2008,Sremcevic-etal-2007}. They
are most probably caused by the presence of moonlets about hundred
meters in size, embedded in the ring, and scattering ring particles
\citep{SpahnSremcevic2000}. As they are embedded in the ring, these
small bodies should exchange angular momentum with the ring, and
migrate \citep{Crida-etal-2009DPS}. This migration could be detected
by Cassini observations through the cumulative lag, or advance with
time $t$ of the orbital longitude $\phi$ induced by a small variation
of the semi-major axis and the angular velocity $\Omega$
($\delta\phi=\delta\Omega\times\delta t$), offering for the first time
the possibility to confront directly the planetary migration theory
with observations, and to give insights and constrains on the physical
properties of the rings and of the moonlets.

In this paper, we address the question of the theoretical migration
rate of these propellers, using both numerical and analytical
approaches. The theory is then confronted to observations. In
\Sect{sec:type1}, we review the standard theory of type~I migration\,;
the differences between migration in protoplanetary disks and in
Saturn's rings are explained, showing the need for a new calculation
of the migration rate of embedded moonlets. This rate is given in
\Sect{sec:moonlet} for an homogeneous, axisymmetric disk with a flat
surface density profile, as a result of numerical computation in
\Sect{sub:num}, and analytical calculation in \Sect{sec:john}. In
\Sect{sec:type4}, we consider the effect of density fluctuations in
the rings, in particular the role of short-lived gravitating clumps,
also called gravity wakes, which are known to be numerous in the
A~ring \citep{Colwell-etal-2006}. We then conclude in
\Sect{sec:conclu} on what our model tells us on the properties of the
rings, given the observed migration rates.

\section{Review of Type~I migration and the differential Lindblad torque}
\label{sec:type1}

In protoplanetary gaseous disks, the perturbation caused by a
terrestrial planet leads to the formation of a one armed spiral
density wave, leading the planet in the inner disk, and trailing
behind the planet in the outer disk. This wave is pressure supported
and generally called the wake, but it has nothing to do with the
gravity wakes mentioned above\,: the latter are local features, while
the planet wake spirals through the whole disk. The planet wake
carries angular momentum, so that the angular momentum given by the
planet to the disk is not deposited locally \citep[e.g.][Appendix
  C]{Crida-etal-2006}. Therefore, the disk profile is hardly modified
in this linear regime. Still, the negative torque exerted by the outer
disk on the planet through the wake is larger in absolute value than
the positive torque from the inner disk. Without going into the
details \citep[for which the reader is refereed to][]{Ward1997}, the
main reason the outer disk wins over the inner disk lies in pressure
effects\,: from the dispersion equation of a pressure supported wave,
one finds that the location $r_{L,m}$ where the wave with azimuthal
mode number $m$, corresponding to the $m$th Lindblad resonance with
the planet, is launched, is not exactly the location of the resonance
given by Kepler's laws. The shift is not symmetrical with respect to
the planet position for inner and outer resonances, but favors the
outer ones. As a consequence, the planet feels a negative total
torque, called the {\it differential Lindblad toque}, and given by
\citet{Tanaka-etal-2002}\,:
\begin{equation}
T_{\rm diff} = -C q^2 \Sigma {r_p}^4 {\Omega_p}^2 h^{-2}\ ,
\label{eq:typeI}
\end{equation}
where the index $p$ refers to the planet, $r_p$ being the radius of
its orbit and $\Omega_p$ its angular velocity, $q$ is the
planet to primary mass ratio, $\Sigma$ is the surface density of the
disk in the neighborhood of the planetary orbit. Finally, $h$ is
the aspect ratio of the disk, being the ratio between its scale height
and the distance to the central body, being proportional to the
square root of the gas pressure. In a typical protoplanetary disk,
$h\approx 0.05$. The numerical coefficient $C$ is given by $C =
2.340-0.099\xi$, where $\xi$ is the index of the power law of the
density profile\,: $\Sigma \propto r^{-\xi}$.

This result is robust. In particular, the value of the negative torque
is almost independent on the slope of the density profile $\xi$. This
is due to the so-called {\it pressure buffer}\,: the resonances are
shifted when the density gradient varies \citep{Ward1997}. If the
disk were pressureless, then the expression of $C$ would be completely
different. Also, the aspect ratio $h$ in \Eq{eq:typeI} appears because
$r_{L,m}$ doesn't converge towards $r_p$ when $m$ tends to infinity,
but towards $r_p(1 \pm 2h/3)$, due to pressure effects. To sum up, the
gas pressure plays a fundamental role in type~I migration.

It should be mentioned for completeness that, in addition to the
differential Lindblad torque discussed above, the horseshoe drag
--\,exerted on the planet by the gas on horseshoe orbits around the
planetary orbit\,-- plays a significant role in type~I migration
\citep[see e.g.][]{Ward1991,Masset2001,BaruteauMasset2008AD,
KleyCrida2008,Paardekooper-Papaloizou-2009a,Paardekooper-etal-2009}.

In contrast to gaseous protoplanetary disks, pressure effects in
Saturn's rings are not important. The aspect ratio $h$ is of the order
of $10^{-7}$. The spiral density waves that are observed in the A ring
are gravity supported, not pressure supported. Thus, the standard
theory of type~I migration does not apply. In particular, the spiral
planet wake doesn't appear. The interaction of the moonlet responsible
for the propeller structure with the disk is observed to take place
within a few hundred kilometers. Resonances with $m\gtrsim 10^3$ are
located within this distance and should play a significant
role. However, in the standard type~I migration the important
resonances have $m \sim 1/h \sim 10^7$.  Therefore, \Eq{eq:typeI}
can't be directly applied to a moonlet in Saturn's rings. A new
approach is needed, adapted to the two main characteristics of the
problem\,: the fact that the rings are made of solid particles, and
the fact that the interaction is taking place very close to the
moonlet.

\section{The ring-moonlet interaction}
\label{sec:moonlet}

In this section, we compute the interaction between a moonlet and a
ring test particle. In this analysis, the gravity of the other ring
particles is neglected. This leads to the torque exerted on the
moonlet by an initially unperturbed, homogeneous ring. In
subsection~\ref{sub:num}, the computation is performed numerically. In
subsection~\ref{sec:john}, it is derived analytically. Both results
are in agreement, and a corresponding migration rate for the moonlet
is given and discussed in subsection~\ref{sub:migr}.

From now on, $m$ denotes the mass of the moonlet (and not anymore
the order of a resonance). The moonlet, has a circular orbit of
radius $r_m$ around the central planet of mass $M$. The gravitational
potential due to the moonlet is $\Psi$. The radial and azimuthal
components of the equation of motion of a ring particle in two
dimensional cylindrical polar coordinates $(r,\phi)$ are
\begin{equation}
\frac{d^2 r}{dt^2}-r\left(\frac{d\phi}{dt}\right)^2 = -\frac{\partial \Psi}{\partial r}-\frac{GM}{r^2}
\label{motr}\\
\end{equation}
\begin{equation}
{\rm and} \ \ \ \ \ \ \ \ \ 
r\frac{d^2 \phi}{dt^2}+2\left(\frac{d r}{dt}\right)\left(\frac{d\phi}{dt}\right) = -\frac{1}{r}\frac{\partial \Psi}{\partial \phi}.
\label{motphi}
\end{equation}
The angular velocity of the moonlet is $\omega=\sqrt{GM/{r_m}^3}$. Let
$r_0$ be the radius of the initially circular orbit of a test particle
and $\Omega=\sqrt{GM/{r_0}^3}$ its angular velocity. We note
$b=r_0-r_m$ is the impact parameter, and $\hat{b}=b/r_H$ the
normalized impact parameter, where
$r_H=r_m\left(\frac{m}{3M}\right)^{1/3}$ is the Hill radius of the
moonlet.

\subsection{Numerical computation of the ring moonlet interaction}
\label{sub:num}

\begin{figure}
\begin{center}
\includegraphics[width=0.4\linewidth,angle=270]{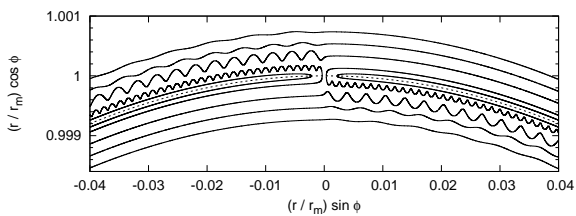}
\end{center}
\caption{Trajectories of test particles perturbed by a moonlet of mass
  $m=3\times 10^{-12}M$, located at $(r=r_m,\phi=0)$ (that is at
  $(0,1)$ in the plot), in the frame corotating with the
  moonlet. Dashed circle\,: orbit of the moonlet.}
\label{fig:traj}
\end{figure}

In this section, the numerical integration of the above equations of
motion is performed, in order to find the trajectories of ring
particles in the presence of a perturbing moonlet in the frame
corotating with the moonlet.  To measure the tiny asymmetry between
the inner and the outer part of the ring, the full equations are
integrated, without linearization or simplification. A Bulirsch-Stoer
algorithm \citep{NumRecipesF1992} is used, and a Taylor expansion is
performed in the code when necessary to subtract accurately large
numbers, in order to achieve machine double precision ($10^{-16}$). We
have checked that the Jacobi constant is conserved to this precision
along the trajectories. Examples of obtained trajectories are given in
\Fig{fig:traj}.

It is well known within the framework of the restricted 3-body problem
that if $|\hat{b}|$ is small enough, the test particle has a horseshoe
shaped orbit, while if $|\hat{b}|$ is larger than $\sim 2.5$, the test
particle is circulating, and scattered into an eccentric orbit. This
can be seen in \Fig{fig:traj}. We perform many numerical integrations
with various $b$ in the case of a moonlet of mass $m=3\times 10^{-12}
M$, starting the particle at an azimuth $|\phi_0|=3000\,r_H/r_m=0.3$
(where the moonlet is at $\phi=0$). This angle is large enough so that
at this location the influence of the moonlet is negligible and the
orbital parameters of the test particle are not disturbed, as will be
checked later. We find that the horseshoe regime occurs for
$\hat{b}<1.8$, and the scattered regime occurs for $\hat{b}>2.5$. For
$1.774<\hat{b}<2.503$, however, the trajectory approaches the center
of the moonlet to within a distance smaller than $0.95\,r_H$. In that
case, if one assumes the moonlet is a point mass, the test particle
eventually leaves the Hill sphere, either on a horseshoe or a
circulating trajectory, but the outcome changes several times with
increasing $\hat{b}$. In the case we are concerned about here, the
moonlet most likely almost fills its Roche lobe, and therefore we stop
the integration of the trajectory as soon as the distance between the
test particle and the moonlet is less than $0.95\,r_H$, assuming a
collision.

\begin{figure}
\includegraphics[width=0.7\linewidth,angle=270]{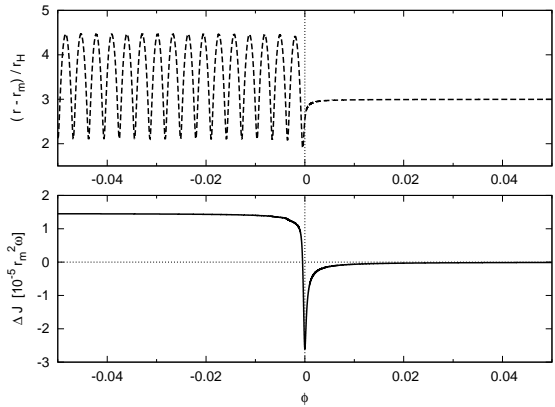}
\caption{Top panel\,: trajectory of a test
  particle with impact parameter $\hat{b}=3$\,; the motion of the
  particle is toward negative $\phi$. Bottom panel\,:
  variation of the specific orbital angular momentum $J$ of the same
  particle along its trajectory.}
\label{fig:Delta}
\end{figure}

The specific orbital angular momentum $J=r^2(d\phi/dt)$ of the test
particles is computed along the trajectories. Angular momentum is
exchanged during the close encounter with the moonlet. For
$\hat{b}\geqslant 2.503$, the test particle is scattered onto an
eccentric orbit of larger angular momentum than initially, which
results in a gain in angular momentum. The variation of orbital
angular momentum along the trajectory is shown in the bottom panel of
\Fig{fig:Delta} for the case $\hat{b}=3$, where the top panel is the
trajectory. The difference in angular momentum between the initial
circular orbit at $\phi_0=0.3\,\rm{sgn}(b)$ and the end of the
integration, when $|\phi|=0.3$ again, is noted $\Delta J$. In the
figure, only the interval $-0.05<\phi<0.05$ is displayed, for
convenience. Most of the exchange of angular momentum occurs when
$|\phi|<0.01$.

\begin{figure}
\includegraphics[width=0.7\linewidth,angle=270]{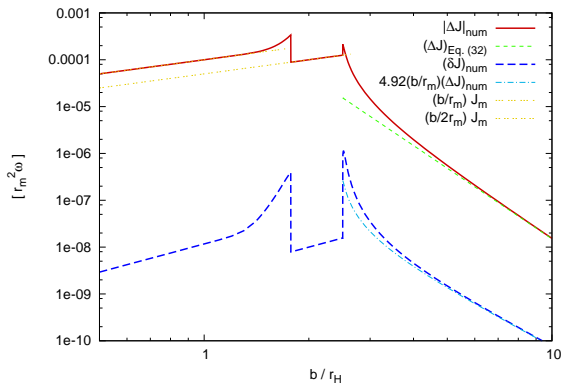}
\caption{Angular momentum exchanges during one close encounter, as a
  function of the impact parameter. Top, thick, red curve\,: $\Delta
  J(b)$, from numerical simulations. Green, thin, dashed, straight
  line\,: $\Delta J(b)$, as given by \Eq{Ampf}. Bottom, thick, dark
  blue, long-dashed curve\,: $\delta J(b)$, from numerical
  simulations. Thin, light blue, dash-dotted line\,: $\delta J(b)$ as
  given by \Eq{Asym}, taking $\Delta J$ from the simulations. Orange,
  thin, double- and triple-dashed lines\,: $(b/r_m)J_m$, and
  $(b/2r_m)J_m$, respectively, to compare with $|\Delta J|$.  }
\label{fig:delta_all}
\end{figure}

Figure~\ref{fig:delta_all} shows $|\Delta J|$ (top thick curve) as a
function of $\hat{b}$, in units of the specific angular momentum of
the moonlet $J_m=r_m^2\omega$. For $0<\hat{b}\leqslant 1.774$, the
horseshoe trajectory corresponds to a U-turn towards the central
planet, and to a loss of angular momentum for the test particle. More
precisely, as for circular orbits $J\propto r^{1/2}$, one expects for
such a U-turn $\Delta J/J = \frac{1}{2}\frac{\Delta r}{r}=-b/r_m$\,;
this is indeed the case for $\hat{b}<1.3$. In the case where the test
particle collides with the moonlet, we assume that it gives all its
orbital angular momentum to the moonlet\,: $\Delta
J=r_0^{\ 2}\Omega-J_m$, so that $\Delta J/J\approx b/(2r_m)$. This
also appears in \Fig{fig:delta_all}. The opposite holds for $b<0$.

Computing $\Delta J$ as a function of $b$ to numerical precision
enables us to also compute the difference between the inner and outer
disk\,: $\delta J(b)=\Delta J(b)+\Delta J(-b)$. This quantity is small
with respect to $\Delta J(b)$, but nonetheless well determined and
converged in our simulations\,: $\Delta J(b)+\Delta J(-b)$ is constant
after the encounter to a precision better than $0.5\%$ for all
$|\phi|>0.02$. This validates our choice of $\phi_0$. In
\Fig{fig:delta_all}, the bottom thick dashed curve shows $\delta J$ in
the same scale as $|\Delta J|$. We see that $\delta J >0$ for all
$b>0$ and that $\delta J \ll \Delta J$, with
\begin{equation}
\delta J/\Delta J \approx 5\times 10^{-4}\,\hat{b}
\label{numAsym}
\end{equation}
for circulating trajectories, and $$\delta J/|\Delta J| \approx
1.17\times 10^{-4}\,\hat{b}$$ for horseshoe orbits.
In the following subsection, the empirically found \Eq{numAsym} is
derived analytically and justified.

\subsection{Analytic model for the ring moonlet interaction}
\label{sec:john}

In this section, we consider only circulating trajectories. Developing
to second order the exchange of angular momentum during an encounter
with the moonlet $\Delta J$, we can find the asymmetry $\delta J$.

\subsubsection{Solution for the perturbed moonlet orbit}

Let us start again from \Eq{motr} and (\ref{motphi}).
The ring particle is assumed to be on an unperturbed circular orbit of
radius $r_0=r_m+b.$ It orbits with angular velocity $\Omega =
\sqrt{GM/{r_0}^3}$ such that $\phi =\Omega t$, where without loss
of generality we have defined the origin of time $t=0$ to be when the
particle is at $\phi=0$. Under the perturbation induced by $\Psi$,
the particle moves to $r=r_0+x,$ and $\phi = \Omega t + y/r_m$, where
$x$ and $y$ are assumed to be small. Linearizing \Eq{motr} and
(\ref{motphi}) about the circular orbit state, we obtain equations for
$x$ and $y$ in the form
\begin{equation}
\frac{d^2 x}{dt^2}-2\Omega \frac{d y}{dt} -3\Omega^2 x=-\left . \frac{\partial \Psi}{\partial r}\right |_0
\  \  \ {\rm and} \  \  \ 
\label{motrp}
\end{equation}
\begin{equation}
\frac{d^2 y}{dt^2}+2\Omega \frac{d x}{dt}=-\left . \frac{1}{r_0}\frac{\partial \Psi}{\partial \phi}\right|_0 
\label{motphip}
\end{equation}
Here the subscript $0$ denotes evaluation on the unperturbed particle
orbit.

We suppose that the perturbation is induced by a moonlet of mass $m$
that is on a circular orbit of radius $r_m$ and has an angular velocity
$\omega.$ Then its azimuthal coordinate $\phi_m=\omega t +\phi_{m,0},$
with $\phi_{m,0}$ being a constant. The perturbing potential
\begin{equation}
\Psi = \frac{Gm}{\sqrt{r_0^2+r_m^2-2r_mr_0\cos(\phi-\phi_m)}}
\label{pot}
\end{equation}
becomes a function of time through substituting $\phi-\phi_m=
(\Omega-\omega)t -\phi_{m,0}$ therein.

Thus we have 
\begin{equation}
\left . \frac{1}{r_0}\frac{\partial \Psi}{\partial \phi}\right|_0 \equiv 
 \left . \frac{1}{r_0(\Omega-\omega)}\frac{\partial \Psi}{\partial t}\right|_0.
\label{mo}
\end{equation}
Using this in \Eq{motphip} and integrating with respect to time, we
obtain
\begin{equation}
\frac{d y}{dt}+2\Omega  x =
 -\left . \frac{\Psi}{r_0(\Omega-\omega)}\right|_0,
\label{motphipi}
\end{equation}
which when combined with \Eq{motrp} gives an equation for $x$ in
the form
\begin{equation}
\frac{d^2 x}{dt^2} + \Omega^2x = -\left .\left( \frac{\partial \Psi}{\partial r} 
+\frac{2\Omega \Psi}{r_0(\Omega-\omega)}\right) \right |_0=S
\label{linearreq}
\end{equation}

\subsubsection{Solution of the linearized equations}

To solve \Eq{linearreq}, we note that the perturbing potential
\Eq{pot} evaluated on the unperturbed orbits is a periodic function
of time with period $2\pi/|\omega-\Omega|=2\pi/\beta.$ Thus we should
look for a periodic response. In order to do this we have to introduce
a small frictional term into \Eq{linearreq} to enable transients to
decay and a net torque on the moonlet to be set up.  When the
frictional term is small it is expected that the resulting torque
should not depend on it \citep[e.g.][]{GT80}. Hence we add
a frictional term $\gamma (dx/dt)$ to the left hand side of
\Eq{linearreq}, where $\gamma/\Omega$ is a small constant parameter
so that it now reads
\begin{equation}
\frac{d^2 x}{dt^2} + \gamma \frac{dx}{dt} +\Omega^2 x=-\left .\left( \frac{\partial \Psi}{\partial r} +\frac{2\Omega\Psi}{r_0(\Omega-\omega)}\right) \right |_0=S\ .
\label{linearreqdamp}
\end{equation}
As the potential is periodic in time we can adopt a Fourier series of
the form
\begin{equation}
S = \sum _{n=-\infty}^{n=\infty}S_n\exp(in\beta t)\ ,
\label{Fourierpot}
\end{equation}
where it is implicit that the real parts of such complex expressions
is to be taken, and
\begin{equation}
S_n = \frac{1}{2\pi }\int_0^{2\pi/\beta}  S(t)\exp(-in\beta t)dt\ .
\label{Fourierpotcoef}
\end{equation}
The periodic solution of \Eq{linearreqdamp} is now readily written
down as
\begin{equation}
x = \sum _{n=-\infty}^{n=\infty}\frac{S_n\exp(in\beta t)}{(\Omega^2-n^2\beta^2+i\gamma n \beta)}.
\label{xFourier}
\end{equation}
We may write this in terms of a Green's function defined through
\begin{equation}
G(\tau)  =\frac{1}{2\pi}
{ \sum _{n=-\infty}^{n=\infty}\frac{\exp(in\beta \tau)}{(\Omega^2-n^2\beta^2+i\gamma n \beta)}}.   
\label{FourierGrn}
\end{equation}
Then the solution for $x$ may be written
\begin{equation}
x =\beta \int_0^{2\pi/\beta}  S(t-t') G(t')dt'.
\label{Grnfsol}
\end{equation}
Note that as the orbit of a ring particle relative to the planet is
periodic, the solution given by \Eq{Grnfsol} includes the effects
of infinite numbers of repeating encounters.  However, we wish to
consider the case when dissipative effects, although weak, are strong
enough to recircularize orbits between encounters in which case they
will be independent of each other. This condition requires that
$\gamma/|\omega-\Omega|=\gamma/\beta \gg 1.$ This is equivalent to
requiring that the damping time scale be short compared to the
relative orbital period between moonlet and ring particle.  On account
of the length scale of the encounters of interest being comparable to
the Hill radius of the moonlet, this is much longer than the orbital
period itself, so that we may adopt the ordering
\begin{equation}
\gamma/|\omega-\Omega|=\gamma/\beta \gg 1 \gg \gamma/\Omega.
\label{inequ}
\end{equation}
In order to make use of the above ordering we write down the form of
the Green's function derived in the appendix (see \Eq{GRNF}\,) valid
for $0 < t < 2\pi/\beta.$
$$G(\tau) = $$
\begin{equation}
\frac{e^{-\gamma\tau/2}\sin(\omega_{\gamma}\tau)-   e^{-\gamma\pi/\beta}\sin(\omega_{\gamma}(\tau-2\pi/\beta))} {\omega_{\gamma}\beta \left[1+e^{-2\gamma\pi/\beta}-2e^{-\gamma\pi/\beta}\cos(2\pi\omega_{\gamma}/\beta)\right]}\ ,
\label{GRNFa}
\end{equation}
where $\omega_{\gamma}=\sqrt{\Omega^2-\gamma^2/4}.$ The function is
defined elsewhere through its periodicity with period $2\pi/\beta.$
Making use of the inequality \Eq{inequ} we may replace the Green's
function \Eq{GRNFa} by the simple expression
\begin{equation}
 G(\tau) =  \frac{\exp(-\gamma\tau/2)\sin(\omega_{\gamma}\tau)}{\omega_{\gamma}\beta}\ .
\label{GRNFAPPROX}
\end{equation}
Then the solution \Eq{Grnfsol} gives 
\begin{equation}
 x = A(t) e^{-\gamma t/2}\sin(\omega_{\gamma} t)    +   B(t) e^{-\gamma t/2}\cos(\omega_{\gamma} t),
\label{episol}
\end{equation}
where
\begin{equation}
  A(t)=  \frac{1}{\omega_{\gamma}}\int^t_{t-2\pi/\beta}S(t') \exp(\gamma t'/2) \cos(\omega_{\gamma} t')dt'
\label{Aamp}
\end{equation}
and
\begin{equation}
  B(t)= - \frac{1}{\omega_{\gamma}}\int^t_{t-2\pi/\beta}S(t') \exp(\gamma t'/2) \sin(\omega_{\gamma} t')dt' .
\label{Bamp}
\end{equation}

To make use of the above expressions, we consider the situation when
the ring particle has a close encounter with the moonlet at time
$t=0,$ thus we take $\phi_{m.0}=0$ (we note that a non zero
$\phi_{m,0}$ can be dealt with by rotating the coordinate system and
shifting the origin of time). The source term $S$ is then expected to
be highly peaked around $t'=0,$ and almost all of the contributions to
the above integrals will occur for $|t'| < \sim 2\pi/\Omega.$
Furthermore, during this dynamical interaction, dissipation will be
negligible. Thus, if we are interested in times after the main
interaction, but before significant dissipation takes place, we may
set $\gamma=0$ and extend the limits of the integration to
$\pm\infty. $ However, in practice one may have to apply a cut off to
the potential at large distances from the moonlet in order to do that
(see below). But this should not matter if the important interaction
occurs when the moonlet and ring particle are close.

Then we simply have
\begin{equation}
  A =   \frac{1}{\Omega}\int^{\infty}_{-\infty}S(t')  \cos(\Omega t') dt'
\label{Aamp1}
\end{equation}
and
\begin{equation}
  B = - \frac{1}{\Omega}\int^{\infty} _{-\infty}S(t')  \sin(\Omega t') dt' .
\label{Bamp1}
\end{equation}
Thus $A$ and $B$ are constants representing epicyclic oscillation
amplitudes induced after the close approach of the ring particle to
the moonlet.

We remark that the approximations made in obtaining \Eq{Aamp1} and
\Eq{Bamp1} relate to how dissipation is treated. There has been no
assumption that the particle trajectories are symmetric on opposite
sides of the moonlet so that curvature effects remain fully
incorporated during particle moonlet encounters.  When dissipation is
negligible during the encounter, then immediately afterward an
epicyclic oscillation is established.  The assumption that dissipation
circularizes orbits between encounters implies that we should consider
approaching ring particles to be on circular orbits.  The above
discussion indicates that errors associated with this assumption are
exponentially small.

\vfill

\subsubsection{Angular momentum transfer}

For the set up considered here, symmetry considerations imply that
$S(t)$ is an even function of time (see below), so that $B=0$. The
generation of the epicyclic oscillation is associated with an angular
momentum transfer between the moonlet and particle. To find this we
firstly note that
\begin{equation}
\Delta J = \sqrt{GM}\left( \sqrt{a_f(1-e^2)}-\sqrt{r_0}\right),
\label{DeltaJb}
\end{equation}
where $a_f$ and $e$ are the post encounter semi-major axis and
eccentricity of the particle. We also note that the Jacobi constant
implies that the change of the particle orbital energy and angular
momentum are related by $\Delta E= GM[1/(2r_0)-1/(2a_f) ]=
\omega\,\Delta J$. This can be used to eliminate $a_f$ in \Eq{DeltaJb}
after which $\Delta J$ may be found correct to second order in $e
\equiv A/r_0$ with the result that $\Delta J = \Omega^2A^2/
[2(\omega-\Omega)]$. This in turn may be simply determined after
evaluating $A$. Note that $\Delta J < 0 $ for particles interior to
the moonlet which have $\Omega>\omega$ and conversely $\Delta J > 0$
for particles orbiting exterior to the moonlet.

\vfill

\subsubsection{Development of the perturbing potential}

We now consider
\begin{equation}
S(t) =-\left .\left( \frac{\partial \Psi}{\partial r} 
+\frac{2\Omega\Psi}{r_0(\Omega-\omega)}\right) \right |_0.
\label{Src}
\end{equation}
\newpage
We begin by recalling that
$$\Psi= - \frac{Gm}{\sqrt{r_0^2+r_m^2-2r_0r_m\cos(\phi-\phi_m)}}$$
\begin{equation}
    = - \frac{Gm}{\sqrt{r_0^2+r_m^2-2r_0r_m\cos(\beta t)}}.
\label{linearpot}
\end{equation}
In order to evaluate the Fourier transform as specified by
\Eq{Aamp1}, which was derived under the assumption that the
interaction occurs only near closest approach, we must truncate the
potential at large $|t|.$ As the encounter takes place over a time
$\ll 1/\beta,$ this can be achieved by replacing $\cos(\beta t)$ in
\Eq{linearpot} by $1- \beta^2 t^2/2.$ Note that a dimensionless
estimate of the error involved is of order $(\beta/\omega)^2 \sim
(r_H/r_m)^2,$ where $r_H$ is the Hill radius of the moonlet.  This is
small enough that the leading order asymmetry in the angular momentum
transferred to orbits with the same impact parameter on either side of
the disk can be estimated.

As the first stage in evaluating the Fourier transform of $S$
specified in \Eq{Aamp1} that gives the epicyclic amplitude, we
evaluate
$$ C = \frac{1}{\Omega}\int^{\infty}_{-\infty}  \Psi \cos(\Omega t)dt \hspace{3cm} $$
\begin{equation}
 = - \frac{1}{\Omega}\int ^{\infty}_{-\infty}   \frac{Gm\cos(\phi) }
{\frac{\beta}{\Omega}\sqrt{(r_0-r_m)^2\Omega^2/\beta^2+r_0r_m\phi^2}}\frac{d\phi}{\Omega}\ .
\label{potcoef}
\end{equation}
This can also be expressed as
\begin{equation}
  C = -\frac{2 GmK_0(\xi_0)}{\Omega\beta \sqrt{r_0r_m}},
\label{K0}
\end{equation}
where $\xi_0 = (\Omega |r_0-r_m|)/(\beta\sqrt{r_0r_m}),$ and $K_j$
denotes the modified Bessel function of the second kind of order $j.$

\subsubsection{Total angular momentum exchange}

We may now use the above expression together with \Eq{Src} to
evaluate the epicyclic amplitude \Eq{Aamp1} (noting that the radial
derivative is with respect to $r_0$ with other quantities held fixed)
so obtaining
$$  A = -\frac{2 Gm}{\Omega\beta r_0\sqrt{r_0r_m}} \times\left(K_0(\xi_0)\left[\frac{1}{2}-\frac{2\Omega}{(\Omega-\omega)}\right]\right. $$
\begin{equation}
\left. +\ K_1(\xi_0)\xi_0\left[\frac{1}{2} +\frac{r_m}{(r_0-r_m)}\right]\right).
\label{Amp1}
\end{equation}
The associated angular momentum exchanged is then given by
$$  \Delta J =  \frac{2 (Gm)^2}{r_0^3r_m(\omega-\Omega)^3} \times \left(K_0(\xi_0)\left[\frac{1}{2}-\frac{2\Omega}{(\Omega-\omega)}\right]\right.$$
\begin{equation}
\hfill \left. + K_1(\xi_0)\xi_0\left[\frac{1}{2} +\frac{r_m}{(r_0-r_m)}\right]\right)^2.
\label{DJamp1}
\end{equation}

In a strictly local approximation under which the inner and outer
sides are symmetric, the contributions from orbits equidistant from the
moonlet would cancel, leaving the net result to be determined by the
surface density profile. However, although we have assumed the
interactions are local, we did not assume symmetry between the
exterior and interior orbits. Accordingly we evaluate the difference
in the magnitude of $\Delta J$ evaluated from orbits equidistant from
the moonlet\,: $r_0=r_m\pm b$. The leading order contribution to
$\Delta J$ is symmetric in $b$. The lowest order contribution is
antisymmetric and accordingly leads to cancellation between the two
sides. We make use of the expansions $\xi_0 =2/3-b/(2r_m)
+O((b/r_m)^2)$, and $2\Omega/(\Omega-\omega)=-4r_m/(3b)(1-b/(4r_m))
+O(b/r_m)$ together with standard properties of Bessel functions to
write

\begin{equation}
  \Delta J =  \frac{64 (Gm)^2r_m}{243\omega^3b^5}
  \left(2K_0(2/3)+K_1(2/3)\right)^2
\left(1+\alpha\frac{b}{r_m} \right),
\label{Ampf}
\end{equation}
where 
\begin{equation}
  \alpha =  \frac{3}{4}+
  \frac{(6K_1(2/3)+3K_0(2/3))}{(4K_0(2/3)+2K_1(2/3)}=2.46\ .
\label{alphaAmpf}
\end{equation}
The first order term of \Eq{Ampf} was already given by
\citet{GT80}. It is plotted as a straight green dashed line in
\Fig{fig:delta_all}. Our expansion to second order enables us to go
further, and to give the expression of the magnitude of the asymmetry
between the two sides of the disk\,:
\begin{equation}
\frac{\delta J}{\Delta J} = 2 \alpha|b|/r_m  = 4.92|b|/r_m\ .
\label{Asym}
\end{equation}
It is such that for an orbit with a given impact parameter, the
angular momentum exchanged in the outer disk is the larger.

In the case studied numerically, we had $r_H=10^{-4}$, so that
$|b|/r_m=10^{-4}\hat{b}$. Then, \Eq{Asym} remarkably agrees with the
numerical fit \Eq{numAsym}. The light blue dot-dashed curve in
\Fig{fig:delta_all} displays $4.92\times 10^{-4}\,\hat{b}\,\Delta J$.

In the context of the above, we note that approximations made in
obtaining equation \Eq{DJamp1} such as effectively starting and
truncating the interaction at some finite though large distance from
the moonlet could conceivably lead to changes comparable to those
given by \Eq{Asym}. However, such changes are again approximately
symmetric for trajectories on both sides of the moonlet and thus
approximately cancel so we do not expect such effects to significantly
alter \Eq{Asym}.

\subsection{Migration rate and discussion}
\label{sub:migr}

\begin{figure}
\includegraphics[width=0.7\linewidth,angle=270]{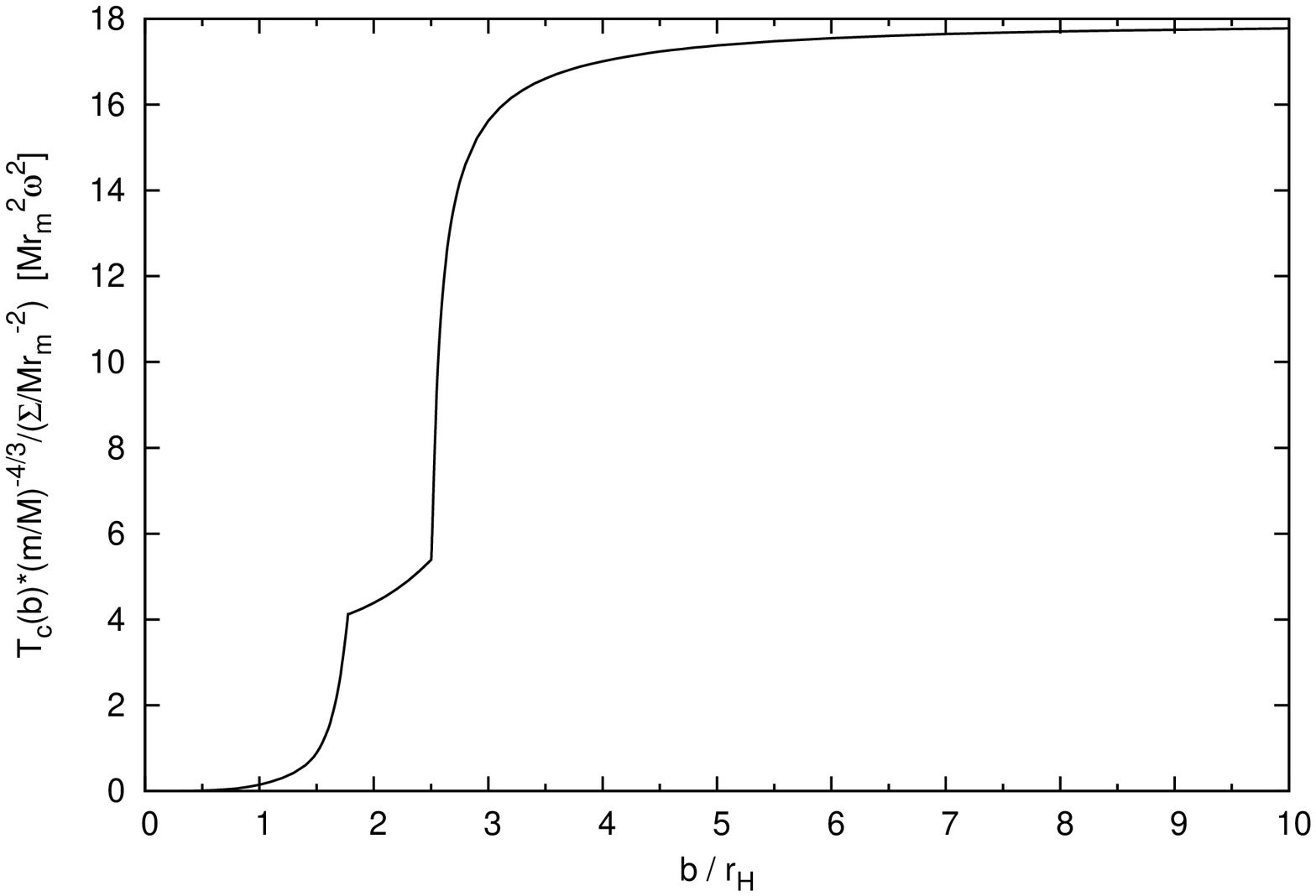}
\caption{Cumulative torque given by \Eq{eq:cumtorque}, exerted by
  a moonlet on the region of the ring $r_m-b<r<r_m+b$.}
\label{fig:cumul}
\end{figure}

If the surface density of ring particles is $\Sigma$, the total rate
of angular momentum transferred to the moonlet is
\begin{equation}
\frac{dJ}{dt}= -\int\!\!\int_{\rm disk}\Sigma \ \Delta J\, \frac{|\omega-\Omega|}{2\pi}dr\,rd\phi\ ,
\end{equation}
where the integral is taken over the disk. The particles exterior to
the moonlet contribute negatively while those interior contribute
positively. The cumulative torque exerted
by the moonlet on the region of the ring located within a distance $b$
to its orbit reads then\,:
\begin{equation}
T_c(b) = \int_{-b}^b \Sigma (r_m+b') (\Delta J(b')) |\omega-\Omega| db'
\label{eq:cumtorque}
\end{equation}
The normalized cumulative torque
$$T_c(b)/\left[(m/M)^{4/3}(\Sigma/Mr_m^{\,-2})\right]$$ is plotted in
\Fig{fig:cumul}. The proportionality to $\Sigma$ is obvious\,; that
$T_c\propto(m/M)^{4/3}$ is numerically verified for $3\times 10^{-15}
\leqslant m/M \leqslant 3\times 10^{-9}$, and has been already found
analytically by \citet{Ward1991} for the horseshoe drag in a similar
context.

Most of the total torque comes from scattered, circulating particles,
in particular the ones with smallest impact parameter $\hat{b}\approx
2.5$. This makes the total torque sensitive to the physical size of
the moonlet (taken as $0.95\,r_H$ here), as some particles colliding
with the moonlet could be circulating if it were smaller.

The role of the horseshoe drag appears to be non negligible, amounting
to 

\noindent $\sim 4.1 (\Sigma/Mr_m^{\,-2})(m/M)^{4/3}\ Mr_m^{\ 2}\omega^2$. The
expression of \citet{Ward1991} for the torque arising from material
executing horseshoe turns, called the horseshoe drag, is for a
Keplerian disk with flat density profile\,:
\begin{equation}
T_{\rm HS} = \frac{9}{8}\Sigma{w}^4\omega^2\ ,
\label{eq:Ward}
\end{equation}
where $w$ is the half-width of the horseshoe region. In our case,
$w=1.774\,r_H$, which gives $T_{\rm HS} =
2.6\,(\Sigma/Mr_m^{\,-2})(m/M)^{4/3}\ Mr_m^{\ 2}\omega^2$. The
agreement is good because Ward's analysis is based only on geometrical
effects and angular momentum variation in a Keplerian disk, without
any pressure effect. Therefore, it also applies in Saturn's ring. We
remark that taking $w=2r_H$ in \Eq{eq:Ward} gives a perfect match with
what we find numerically for the total horseshoe drag.

In conclusion, from \Fig{fig:cumul}, the total torque felt by a
moonlet of mass $m$ on a circular orbit of radius $r_m$ around a
planet of mass $M$ can be written as
\begin{equation}
T = - 17.8\,\left(\frac{\Sigma}{Mr_m^{\,-2}}\right)\left(\frac{m}{M}\right)^{4/3}\ Mr_m^{\ 2}\omega^2\ .
\end{equation}
Note that to get the same dependency of the type~I torque in the
parameters of the system, one has to assume $h\propto r_H/r_m$ in
\Eq{eq:typeI}\,; however, in a protoplanetary disk, $h$ is fixed and
independent of the mass of the secondary body, so that this
proportionality would not be justified.

The torque is related to the migration speed through
  $T=0.5\,m\,r_m\Omega(dr_m/dt).$ Hence we deduce that
\begin{equation}
\frac{dr_m}{dt} = - 35.6\,\frac{\Sigma r_m^{\ 2}}{M}\left(\frac{m}{M}\right)^{1/3}\ r_m\Omega\ .
\end{equation}
The migration rate is here proportional to the mass of the moonlet to
the power $1/3$, in contrast to standard type~I migration where
$dr_m/dt\propto m$. A numerical application to the case of an $m=10^{-18}
M_{\rm Saturn} = 5.68\times 10^{8}$\,kg moonlet in orbit in the A-ring
of density $\Sigma=400$ kg\,m$^{-2}$ at $r_m=130\,000$~km from Saturn
gives $dr_m/dt=-0.23$~m\,yr$^{-1}$. Increasing the mass by two orders
of magnitude to correspond to a radius of $\sim 200$~m speeds up the
migration rate by a factor of only $\sim 4.5$ to $\sim -1$~m\,yr$^{-1}$.

After time $t$, a migrating propeller will be shifted longitudinally
with respect to a corresponding non migrating one by a distance
$r_m\,\delta \phi = 3\Omega\,|d r_m/dt|\,t^2/4$. For the above
parameters, this gives $713\,[t/(1\,{\rm year})]^2$~m.
A shift of this magnitude is potentially detectable on a timescale of
a year to a few years (\citealp{ Porco-etal-2004} and note also
\citealp{Burns-etal-2009DPS}). Actually, migration of propellers has
already been detected
\citep{Burns-etal-2009DPS,Tiscareno-etal-2010}. During one time period
of nearly a year, a particular propeller has been seen moving {\it
  outward} at a rate of $\sim 110$~m\,yr$^{-1}$\,; and during a later
similar time period, the same propeller has been seen moving {\it
  inward} at a rate of $\sim 40$~m\,yr$^{-1}$ \citep[][and personal
  communication]{Tiscareno-etal-2010}.

These observations are not compatible with the above theory. But we
recall that the process of migration of a moonlet described above
assumed a smooth particle disk with constant surface density. Here we
note that there are features and mechanisms that might produce a
significantly faster migration rate, possibly in both directions
inward and outward, and non constant in time. One can first think of a radial
density gradient\,: as there is no pressure buffer here, this would
directly affect the balance between the torques from the inner and
outer parts of the ring. This would also affect the torque from the
horseshoe region, which could turn positive. However, if the migration
is governed by the gradient of some quantity, it seems likely that the
moonlet would have approached an extremum in that quantity, and thus
should have attained a migration rate comparable to that estimated for
a constant surface density.

Another possibility resulting in the moonlet migrating faster than
what the previous calculation indicates, and possibly outward, is a
runaway migration in a planetesimals disk \citep[][for a
  review]{Ida-etal-2000, Levison-etal-2007_PPV}, similar to the
type~III migration of planets in protoplanetary disks
\citep{MassetPapaloizou2003}. In this regime, the migration of the
moonlet in the disk leads to a positive feedback on its migration
rate, because of the material of the inner (resp. outer) disk making
horseshoe U-turns to the outer (resp. inner) disk. This speeds up the
migration, possibly leading to a runaway. However, this leads
inevitably to an asymmetry in the horseshoe region, while the
propeller structures observed are rather symmetrical.

Finally, the A-ring of Saturn is not homogeneous. It is close to
gravitational instability, which should lead to the formation of
gravity wakes and density fluctuations. The effect of these density
fluctuations on the moonlet is studied in next section.

\section{The role of density fluctuations and resulting stochastic migration}
\label{sec:type4}

The analytic calculations and numerical simulations in the previous chapters assume an inflow of particles on circular orbits only perturbed by the nearby moonlet. However, we know that Saturn's A ring is marginally gravitationally stable \citep{Daisaka2001}. The Toomre $Q$ parameter \citep{Toomre1964}, which is a measure of the importance of self-gravity, is expected to be of the order of $2\sim7$, indicating that the ring particles' mutual gravity is indeed a strong effect. It leads to the regular formation and dispersion of gravity wakes, which are local density enhancements elongated in parallel directions by the Keplerian shear. Those over-densities give rise to stochastic forces which act on the embedded moonlet. 

A very similar effect is expected to occur in protoplanetary disks. These disks are thought to be turbulent due to the magneto-rotational instability \citep[MRI,][]{BalbusHawley1991}. The turbulent fluctuations create over-densities which interact gravitationally with embedded small mass planets. The stochastic forces make the planet undergo a random walk.  An analytic model of this random walk has been derived by \cite{Rein-Papaloizou-2009}. In the following, we apply this model to moonlets embedded in Saturn's rings. To do that, we need to get an estimate of the amplitude of the stochastic forces. 

\subsection{Numerical calculations}
We perform three-dimensional simulations of ring particles, in a shearing box, similarly to \citet{Salo1995}. The simulations are done in a local cube of size $H$ with shear periodic boundary conditions, and the origin of the box is fixed at a semi major axis of $a=130\,000$~km. A BH tree code \citep{Barnes1986} is used to calculate the self-gravity between ring particles and resolve inelastic collisions. Collisions between particles are resolved using the instantaneous collision model and a velocity dependent coefficient of restitution given by \cite{Bridges-etal-1984}\,:
\begin{equation}
\epsilon(v) = \mathrm{min}\left\{0.34\times \left( \frac{v}{1\,\mathrm{cm.s}^{-1}} \right)^{-0.234}, 1\right\},
\end{equation}
where $v$ is the impact speed projected on the vector joining the centers of the two particles. The code is described in more detail in \cite{ReinLesurLeinhardt2010}.

The size of ring particles is not well constrained. Therefore and to be able to scale to different locations in Saturn's rings, we perform multiple simulations. For a given simulation, all the particles are assumed to be spherical and have the same size (or radius), which varies from simulation to simulation from $0.52$~to~$13$~meters. The simulation parameters are listed in Table~\ref{tab:sim}. The nomenclature and physical parameters are, for easy comparison, the same as in \cite{LewisStewart2009}, as our simulations are similar to theirs.

\begin{table*}[htbp]
\begin{center}
\begin{tabular}{l|llcccr}
Name &	$r_a$ &	$\tau$ &   $\rho_p$      &   $\Sigma_p$     &  $H$   & $N$ \\
\hline\hline
L2   & 	 13 m &  0.1  & 0.5 g\,cm$^{-3}$ & 885 kg\,m$^{-2}$  & 5000 m & 4\,808 \\
S1   & 0.52 m &  0.1  & 0.7 g\,cm$^{-3}$ & 49.7 kg\,m$^{-2}$ & 1000 m & 120\,548\\
S3   & 	1.3 m &  0.2  & 0.7 g\,cm$^{-3}$ & 246 kg\,m$^{-2}$  & 1000 m & 38\,188	
\end{tabular}
\caption{Simulation parameters\label{tab:sim}. The first column identifies the simulation, following the convention of \cite{LewisStewart2009}. The second and third column give the size (or radius) of the particles and their density. The fourth and fifth column give the surface density and the size of the computational domain, respectively. The last column lists the number of particles.}
\end{center}
\end{table*}

The moonlet is not taken into account in the simulations. We measure the specific gravitational force $\hat\mathbf{f}$ (or acceleration) felt by a passive test-particle sitting at the origin. We calculate the force in two different ways, in order to avoid the singularity at the origin and to account for the physical size of the moonlet. In the first case, we use a cut off at the moonlet's radius $d$ and exclude all particles within that radius from the force calculation. In the second case we use a smoothed gravitational force per unit mass in the form
\begin{equation}
\hat\mathbf{f} = -\frac{Gm_{\rm part}}{|\hat\mathbf{r}|^2+d^2} \hat\mathbf{r},
\label{eq:acc}
\end{equation} 
where $\hat\mathbf{r}$ is the vector linking the origin to the particle and $m_{\rm part}$ is the mass of the particle. The smoothing length $d$ is set equal to the moonlet's size. In a self-consistent simulation, one should include the moonlet with it's real physical size. However, this goes beyond the scope of this paper and will be considered in future work \citep{Rein-Papaloizou-2010}. Our purpose here is to estimate the underlying stochastic fluctuations in the migration rate that occur independently of the moonlet. This procedure is reasonable as long as the moonlet is in a steady state, namely if it doesn't accumulate or lose a large amount of mass over one orbit. In all our simulations we assume a moonlet size of $d=200$~m.

\subsection{Results}
\begin{figure}
\centering
\includegraphics[width=0.35\textwidth,angle=270]{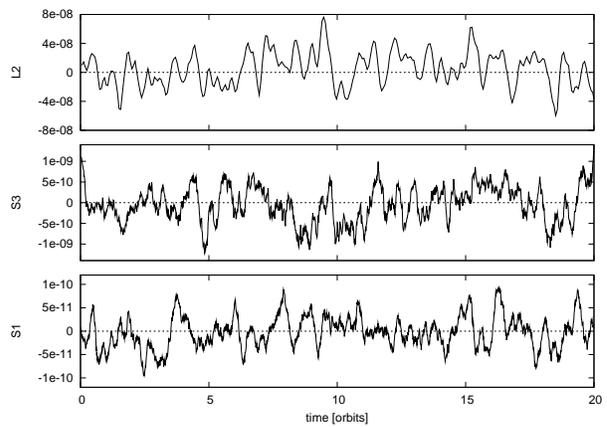}
\caption{Azimuthal component of the specific gravitational force felt by the moonlet in m\,s$^{-2}$ for simulations L2, S3 and S1 (from top to bottom).
\label{fig:plot_forces}}
\end{figure}

We measure the amplitude and the correlation time of the stochastic forces in all simulations. The results are listed in Table~\ref{tab:res}. We also plot the time evolution of the azimuthal ($y$) force component in \Fig{fig:plot_forces}. Whereas the correlation time in all simulations is almost the same, the amplitude of the stochastic fluctuations varies by almost a factor of $10^3$. The forces in the vertical direction are negligible and not presented here. 
The diffusion coefficient, being a measure of the strength of stochastic forces, is defined as $D=2\langle f^2\rangle \tau$ \citep[see][]{Rein-Papaloizou-2009}, where $\langle f^2\rangle^{1/2}$ and $\tau$ are the root mean square value and the approximate correlation time of the specific stochastic force in one direction.

\begin{table*}[htbp]
\begin{center}
\begin{tabular}{ll|ll|ll|ll|c}
\multicolumn{2}{l|}{Simulation} &    \multicolumn{2}{l|}{Correlation time [s]} & \multicolumn{2}{l|}{Diffusion coefficient [m$^2$\,s$^{-3}$]} & Q\\
& &  \ $\tau_x$ & \ $\tau_y$  		& $D_x$ & $D_y$ & \\\hline\hline
L2 & cutoff 	& \ 5000\  & \ 7000  & $9.61\times10^{-12}$ & $14.97\times10^{-12}$&4.1\\
   & smooth 	& \ 6000\  & \ 9000  & $5.34\times10^{-12}$ & $9.11\times10^{-12}$&\\\hline
S1 & cutoff	& \ 2000\  & \ 4000  & $1.92\times10^{-17}$ & $2.26\times10^{-17}$&7.2\\
   & smooth	& \ 3000\  & \ 8000  & $1.59\times10^{-17}$ & $1.69\times10^{-17}$&\\\hline
S3 & cutoff	& \ 2000\  & \ 6000  & $7.50\times10^{-16}$ & $30.24\times10^{-16}$&2.5\\
   & smooth	& \ 4000\  & 10000 & $8.87\times10^{-16}$ & $20.48\times10^{-16}$&\\
 
\end{tabular}
\caption{Simulation results\label{tab:res}. The first column gives the name of the simulation, as defined in Table~\ref{tab:sim}. The second and third columns give the correlation time in the $x$ (radial) and $y$ (azimuthal) direction, respectively. The fourth and fifth columns list the diffusion coefficients. The sixth column is the Toomre $Q$ parameter, as measured in the simulation.  }
\end{center}
\end{table*}

The change in semi major axis $a$ due to the effect of stochastic forces with diffusion coefficient $D$ after time $t$ is given by
\begin{eqnarray}
\Delta a &=& \frac{2}{\omega} \sqrt{Dt},
\label{eq:diff}
\end{eqnarray}
where $\omega$ is the mean motion of the moonlet \citep{Rein-Papaloizou-2009}. 
Note that in this regime, the acceleration of the moonlet doesn't depend on its mass (as can be seen in \Eq{eq:acc}). Therefore, the migration rate and the diffusion coefficient are independent of the mass of the moonlet\,; this might be an observational indication for this migration regime.

Assuming an initial semi-major axis of $a=130\,000$~km and $D\sim 10^{-17}$\,m$^2$\,s$^{-3}$ as found in simulation S1, one can calculate the expected difference in semi major axis after one orbit due to stochastic forces which turns out to be $\Delta a = 0.01$m. For simulation S3, assuming $D\sim 10^{-15}$\,m$^2$\,s$^{-3}$, one finds $\Delta a = 0.11$m. For the simulation L2, assuming $D\sim 10^{-11}$\,m$^2$\,s$^{-3}$, one finds $\Delta a = 10.5$~m. These translate to random walks with standard deviation given as a function of time by $\Delta a = 0.27 \sqrt {t/(1\ {\rm year})}$~m, $\Delta a = 2.7 \sqrt{t/(1\ {\rm year})}$~m, and $\Delta a = 270 \sqrt {t/(1\ {\rm year})}$~m respectively.

\subsection{Discussion}
From the above, it can be seen that increasing the surface density by factors $4-5$ changes the migration rate by two orders of magnitude. Thus, the results show clearly that the surface density $\Sigma$ is much more important than in the regular, type I like migration, presented in \Sect{sec:moonlet}. This can be easily understood with a toy model. The critical unstable wavelength $\lambda$ scales linearly with $\Sigma$ \citep{Toomre1964}. If we assume a fixed moonlet size, the ratio of moonlet size to $\lambda$ therefore changes with $\Sigma$. In the limit where $\lambda$ is much smaller than the moonlet radius, the stochastic forces are negligible as the density distribution is approximately homogeneous on the relevant scales. This is the case in simulation S1. In the other limit where $\lambda$ is larger than the moonlet, the moonlet undergoes a random walk that is similar to that of individual ring particles, as seen in simulation L2. 

The range in migration rates found in simulations shows that over time scales of several years, the migration of a moonlet of  mass $m \sim 10^{-16}M_{\rm Saturn}$ may be dominated by a random walk in some situations (eg. those in simulations S3 and L2, the latter carried out with particles of radius $13$~m). However, in regions of the rings where the surface density is small (eg. simulation S1), the moonlet may be in a regular, non-stochastic migration regime. In that case, the model from \Sect{sec:moonlet} can be applied.

This dependence offers an exciting possibility to constrain the nature of the ring particles and the physical processes occurring in the rings by measuring the migration of moonlets. But note that because regular migration gives a decrease in the semi-major axis that is linear in time, provided it continues to operate, it will always ultimately dominate the behavior  for large time because the spreading of the semi-major axis associated with stochastic migration increases only as the square root of time.

At the present day, the number of observed migration rates is not sufficient to draw a statistically significant conclusion. However, the fact that the migration varies in rate and direction clearly favors the stochastic migration model presented in this section. Considering a migration rate of $|\Delta a|=100$\,m in $t=1$ year in \Eq{eq:diff}, one finds $D=1.4\times 10^{-12}$\,m$^2$\,s$^{-3}$. Taking $|\Delta a|=40$\,m in $t=1$ year gives $D=2.2\times 10^{-13}$\,m$^2$\,s$^{-3}$. This is in the range obtained in the simulations, and tends to favor the case of simulation L2, and $\Sigma \sim 700$~kg\,m$^{-2}$ in the A~ring.

\section{Conclusion}
\label{sec:conclu}

In this paper we have calculated the differential torque exerted on a
moonlet by the outer and the inner disk with a smooth, flat surface
density profile. We performed both an accurate numerical integration
and a second order analytical calculation. These approaches were found
to be in excellent agreement where their domains of validity
overlap. The migration rate found in this case is proportional to the
mass of the moonlet to the power $1/3$. It is about $-1$~m\,yr$^{-1}$
for a $200$~m radius moonlet in the A-ring. This is way too low to
explain the observed migration of the propellers in Saturn's
rings. Nonetheless, density fluctuations in the rings, due to their
proximity to gravitational instability, can lead to stochastic torques
on a moonlet, that may dominate on the timescale of the Cassini
mission, to an extent that depends mainly on the surface density of the
rings. These stochastic torques may account for the observations.

The possibility that the migration of propellers is induced by
stochastic processes rather than by a regular type~I like migration is
therefore very exciting\,: this may help to infer the local surface
density, and therefore the size of the ring particles. Indeed, both
quantities are linked through the optical depth, which is
observationally well constrained. Our estimate of $\Sigma \sim
700$~kg\,m$^{-2}$ is equivalent to $\sim 10$~m size particles in the
A~ring for a single-sized population. This is in good agreement with
available estimates from stellar occultations. For the A~ring, Voyager
occultations \citep{Zebker-etal-1985} find that the radius $r_a$ of
particles follows\,: $0.1$~m (assumed) $< r_a < 11$~m, with a power
index $\sim -3$. For the 28 Sgr occultation
\citep{French-Nicholson-2000}, a range $1$m $<r_a<20$m is found, with
a power index between $-2.7$ and $-3$, and an effective size (the
single average size accounting for the fluctuations in photon count)
being about $7$m \citep{Cuzzi-etal-2009}.

Fortunately the Cassini mission has been extended to 2017. In the
meantime, numerous observations of the propellers will hopefully give
a clear picture of their orbital evolution for a period of 10 years,
representing about $10\,000$ orbits. This will allow us to test the
hypothesis presented in this paper, and in particular to check whether
the propellers really are in stochastic migration, whereas first
results seem to favor the stochastic hypothesis.

Among the questions that still need to be addressed is why all the
propellers seem to be gathered in the A ring, in places apparently
devoid of density waves\,? Indeed the propellers seem gathered in a
couple of narrow radial bands of only about 1000~km width
\citep{Tiscareno-etal-2008}. This is especially surprising since the
A~ring is densely populated by numerous density waves launched by the
nearby small moons (Atlas, Prometheus, Pandora, Janus, Epimetheus). Is
there a systematic mechanism that would eject the propellers away from
density waves\,? Or does the present location of propellers just
reflect the initial location of the parent body, assuming that the
propeller population comprises the fragments resulting from the
destruction of an ancient moon orbiting within the rings\,?

If the moonlets really undergo stochastic migration, then \Eq{eq:diff}
may strongly constrain the age of the propellers, which can't be
larger than the time needed to diffuse over $\Delta a >
1000$~km. Unfortunately, as long as $D$ is unknown \Eq{eq:diff}
doesn't provide any useful information. However, considering that
$\Delta a$ is proportional to the square root of the time, and
assuming that a moonlet migrates about $100$~m in $1$ year, one finds
that $\Delta a=1000$~km for $t=100$ million years. Assuming $|\Delta
a| = 40$~m in $1$ year, we find that it requires $\sim 625$ million
years to diffuse over $1000$~km. Note that for $200$~m radius
moonlets, the spreading due to stochastic migration equates to the
contraction of the semi-major axes occurring as a result of smooth,
regular migration after $\sim 10^4$~\,yr, and then $\Delta a \approx
10$~km. Thus it would take about a million years to migrate through
$\Delta a =1000$~km in this case, largely through the action of the
non-stochastic, regular migration process, if that can be assumed to
operate smoothly and simultaneously with the stochastic migration
process. A $100$~m radius moonlet would migrate through only $500$~km
in the same period, so that the smooth, regular migration process
spreads a population of moonlets of various sizes over $1000$~km in
one to two million years. These could be the times since the
catastrophic disruption of a small moon orbiting at $130\,000$~km from
Saturn, that was broken into smaller moonlets by a meteoritic
impact. On the other hand, an estimate of the lifetime of a Pan size
moon ($\sim 14$ km in radius) against the today's cometary flux is
provided by \citet{Dones-etal-2009} and gives a range between
$100$~Myr and $16$~Gyr, depending on the size distribution of
impactors.  Therefore, the recent occurrence of such an event, about
$4$ to $4.5$ billion years after solar system formation, is
possible. Note also that an age of about $100$~Myr is coherent with
some estimates of Saturn's ring age despite of the lack of fully
satisfactory explanation for their origin \citep[see][for a
  review]{Charnoz-etal-2009a}.

We see that the question of propeller's migration is inextricably
linked to the issue of the origin of Saturn's moons embedded in the
rings, which is still a mystery. \citet{Porco-etal-2007} and
\citet{Charnoz-etal-2007} have jointly proposed that small moons
embedded in the rings could be aggregates of material on an initial
shard denser than ice. When destroyed by meteoritic bombardment, these
could release dense chunks of material that could explain the origin
of the propellers. However, the origin of Saturn's ring system is
still a matter of debate
\citep{Harris1984,Charnoz-Morby-etal-2009}. Knowledge of the age of
the propellers could provide important constraints on the age of the
main ring system and its embedded moons, as there are strong
indications that these could have about the same age, provided these
moonlets hide a dense shard
\citep{Charnoz-etal-2007,Porco-etal-2007}. Understanding the migration
rate of the propellers is therefore an important piece of this puzzle.

\acknowledgments

We thank J. Burns for stimulating discussions and the organisers of
the ``Dynamics of Discs and Planets'' workshop at the Isaac Newton
Institute in Cambridge where these took place, as well as M. Tiscareno
for providing us with migration rates. Hanno Rein was supported by an
Isaac Newton Studentship, STFC, and St John's College, Cambridge.

\newpage

\appendix

\section{Evaluation of the Green's function}
\label{app:green}

Here we evaluate the Green's function defined by
\Eq{FourierGrn} as
\begin{equation}
G(\tau)  = \frac{1}{2\pi}
{ \sum _{n=-\infty}^{n=\infty}\frac{\exp(in\beta \tau)}{(\Omega^2-n^2\beta^2+i\gamma n \beta)}}.   
\label{A1}
\end{equation}
To perform the summation we use the general result that if for a
general periodic function
\begin{equation}
g(\tau) = \sum _{n=-\infty}^{n=\infty}b(n)\exp(in\beta t),   \label{A2}
\end{equation}
with period $2\pi/\beta,$ and $b(n)$ being defined as an integrable
function, we set
\begin{equation}
F(\tau) =  \frac{1}{2\pi}\int^{\infty}_{-\infty}b(n)\exp(in\beta \tau)dn ,
\label{A3}
\end{equation}
then
\begin{equation}
g(\tau) =2\pi  \sum _{n=-\infty}^{n=\infty}F(\tau+2\pi n/\beta ).
\label{A4}
\end{equation}

We set 
\begin{equation}
b(n)= \frac{1}{2\pi(\Omega^2-n^2\beta^2+i\gamma n \beta)}.
\label{A5}
\end{equation}
Then the integral \Eq{A3} defining $F(\tau)$ is readily performed
by contour integration with the result that for $t>0,$
\begin{equation}
 F(\tau) =  \frac{\exp(-\gamma\tau/2)\sin(\omega_{\gamma}\tau)}{2\pi\omega_{\gamma}\beta} ,
\label{A6}
\end{equation}
otherwise $F(\tau)=0.$ Here
$\omega_{\gamma}=\sqrt{\Omega^2-\gamma^2/4}.$ Using the above to
evaluate the sum \Eq{A4} as a geometric progression yields
$g(\tau)\equiv G(\tau)$ for $0 < \tau < 2\pi/\beta$ as
\begin{equation}
 G(\tau) =  \frac{\exp(-\gamma\tau/2)\sin(\omega_{\gamma}\tau)-   \exp(-\gamma\pi/\beta)\sin(\omega_{\gamma}(\tau-2\pi/\beta))}
 {\omega_{\gamma}\beta \left[1+\exp(-2\gamma\pi/\beta)-2\exp(-\gamma\pi/\beta)\cos(2\pi\omega_{\gamma}/\beta)\right]},
\label{GRNF}
\end{equation}
the function is determined elsewhere by its periodicity with period
$2\pi/\beta.$


\bibliographystyle{elsarticle-harv}
\bibliography{crida.bib}

\begin{thebibliography}{46}
\expandafter\ifx\csname natexlab\endcsname\relax\def\natexlab#1{#1}\fi
\expandafter\ifx\csname url\endcsname\relax
  \def\url#1{\texttt{#1}}\fi
\expandafter\ifx\csname urlprefix\endcsname\relax\def\urlprefix{URL }\fi

\bibitem[{{Balbus} and {Hawley}(1991)}]{BalbusHawley1991}
{Balbus}, S.~A., {Hawley}, J.~F., Jul. 1991. {A powerful local shear
  instability in weakly magnetized disks. I - Linear analysis. II - Nonlinear
  evolution}. \apj 376, 214--233.

\bibitem[{{Barnes} and {Hut}(1986)}]{Barnes1986}
{Barnes}, J., {Hut}, P., Dec. 1986. {A hierarchical O(N log N)
  force-calculation algorithm}. \nat 324.

\bibitem[{{Baruteau} and {Masset}(2008)}]{BaruteauMasset2008AD}
{Baruteau}, C., {Masset}, F., Jan. 2008. {On the Corotation Torque in a
  Radiatively Inefficient Disk}. \apj 672, 1054--1067.

\bibitem[{{Bridges} et~al.(1984){Bridges}, {Hatzes}, and
  {Lin}}]{Bridges-etal-1984}
{Bridges}, F.~G., {Hatzes}, A., {Lin}, D.~N.~C., May 1984. {Structure,
  stability and evolution of Saturn's rings}. \nat 309, 333--335.

\bibitem[{{Burns} et~al.(2009){Burns}, {Tiscareno}, {Spitale}, {Porco},
  {Cooper}, and {Beurle}}]{Burns-etal-2009DPS}
{Burns}, J.~A., {Tiscareno}, M.~S., {Spitale}, J., {Porco}, C.~C., {Cooper},
  N.~J., {Beurle}, K., Jan. 2009. {Giant Propellers Outside the Encke Gap in
  Saturn's Rings}. In: Bulletin of the American Astronomical Society. Vol.~41
  of Bulletin of the American Astronomical Society. pp. 559--+.

\bibitem[{{Charnoz} et~al.(2007){Charnoz}, {Brahic}, {Thomas}, and
  {Porco}}]{Charnoz-etal-2007}
{Charnoz}, S., {Brahic}, A., {Thomas}, P.~C., {Porco}, C.~C., Dec. 2007. {The
  Equatorial Ridges of Pan and Atlas: Terminal Accretionary Ornaments?} Science
  318, 1622--.

\bibitem[{{Charnoz} et~al.(2009{\natexlab{a}}){Charnoz}, {Dones}, {Esposito},
  {Estrada}, and {Hedman}}]{Charnoz-etal-2009a}
{Charnoz}, S., {Dones}, L., {Esposito}, L.~W., {Estrada}, P.~R., {Hedman},
  M.~M., 2009{\natexlab{a}}. {Origin and Evolution of Saturn's Ring System}.
  In: {Dougherty}, M.~K., {Esposito}, L.~W., {Krimigis}, S.~M. (Eds.), Saturn
  from Cassini-Huygens. pp. 537--573.

\bibitem[{{Charnoz} et~al.(2009{\natexlab{b}}){Charnoz}, {Morbidelli}, {Dones},
  and {Salmon}}]{Charnoz-Morby-etal-2009}
{Charnoz}, S., {Morbidelli}, A., {Dones}, L., {Salmon}, J., Feb.
  2009{\natexlab{b}}. {Did Saturn's rings form during the Late Heavy
  Bombardment?} Icarus 199, 413--428.

\bibitem[{{Colwell} et~al.(2006){Colwell}, {Esposito}, and {Srem{\v
  c}evi{\'c}}}]{Colwell-etal-2006}
{Colwell}, J.~E., {Esposito}, L.~W., {Srem{\v c}evi{\'c}}, M., Apr. 2006.
  {Self-gravity wakes in Saturn's A ring measured by stellar occultations from
  Cassini}. \grl 33, 7201--+.

\bibitem[{{Crida} et~al.(2009){Crida}, {Charnoz}, {Papaloizou}, and
  {Salmon}}]{Crida-etal-2009DPS}
{Crida}, A., {Charnoz}, S., {Papaloizou}, J., {Salmon}, J., Sep. 2009.
  {Satellite And Propeller Migration In Saturn's Rings}. Vol.~41 of AAS/DPS
  Meeting Abstracts. pp. \#18.07--+.

\bibitem[{{Crida} et~al.(2006){Crida}, {Morbidelli}, and
  {Masset}}]{Crida-etal-2006}
{Crida}, A., {Morbidelli}, A., {Masset}, F., Apr. 2006. {On the width and shape
  of gaps in protoplanetary disks}. Icarus 181, 587--604.

\bibitem[{{Cuzzi} et~al.(2009){Cuzzi}, {Clark}, {Filacchione}, {French},
  {Johnson}, {Marouf}, and L.}]{Cuzzi-etal-2009}
{Cuzzi}, J., {Clark}, R., {Filacchione}, G., {French}, R., {Johnson}, R.,
  {Marouf}, E., L., S., 2009. {Ring Particle Composition and Size
  Distribution}. In: {Dougherty}, M.~K., {Esposito}, L.~W., {Krimigis}, S.~M.
  (Eds.), Saturn from Cassini-Huygens. pp. 459--509.

\bibitem[{{Daisaka} et~al.(2001){Daisaka}, {Tanaka}, and {Ida}}]{Daisaka2001}
{Daisaka}, H., {Tanaka}, H., {Ida}, S., Dec. 2001. {Viscosity in a Dense
  Planetary Ring with Self-Gravitating Particles}. Icarus 154, 296--312.

\bibitem[{{Dones} et~al.(2009){Dones}, {Chapman}, {MacKinnon}, {Kirchoff},
  {Neukum}, and {Zahnle}}]{Dones-etal-2009}
{Dones}, L., {Chapman}, C.~R., {MacKinnon}, B., {Kirchoff}, M.~R., {Neukum},
  G., {Zahnle}, K.~J., 2009. {Icy Satellites of Saturn: Impact Cratering and
  Age Determination}. In: {Dougherty}, M.~K., {Esposito}, L.~W., {Krimigis}, T.
  (Eds.), Saturn from Cassini-Huygens. pp. 613--635.

\bibitem[{{French} and {Nicholson}(2000)}]{French-Nicholson-2000}
{French}, R.~G., {Nicholson}, P.~D., Jun. 2000. {Saturn's Rings II. Particle
  sizes inferred from stellar occultation data}. Icarus 145, 502--523.

\bibitem[{{Goldreich} and {Tremaine}(1979)}]{GT79}
{Goldreich}, P., {Tremaine}, S., Nov. 1979. {The excitation of density waves at
  the Lindblad and corotation resonances by an external potential}. \apj 233,
  857--871.

\bibitem[{{Goldreich} and {Tremaine}(1980)}]{GT80}
{Goldreich}, P., {Tremaine}, S., Oct. 1980. {Disk-satellite interactions}. \apj
  241, 425--441.

\bibitem[{{Harris}(1984)}]{Harris1984}
{Harris}, A.~W., 1984. {The origin and evolution of planetary rings}. In:
  {R.~Greenberg \& A.~Brahic} (Ed.), IAU Colloq. 75: Planetary Rings. pp.
  641--659.

\bibitem[{{Ida} et~al.(2000){Ida}, {Bryden}, {Lin}, and
  {Tanaka}}]{Ida-etal-2000}
{Ida}, S., {Bryden}, G., {Lin}, D.~N.~C., {Tanaka}, H., May 2000. {Orbital
  Migration of Neptune and Orbital Distribution of Trans-Neptunian Objects}.
  \apj 534, 428--445.

\bibitem[{{Kley} and {Crida}(2008)}]{KleyCrida2008}
{Kley}, W., {Crida}, A., Aug. 2008. {Migration of protoplanets in radiative
  discs}. \aap 487, L9--L12.

\bibitem[{{Levison} et~al.(2007){Levison}, {Morbidelli}, {Gomes}, and
  {Backman}}]{Levison-etal-2007_PPV}
{Levison}, H.~F., {Morbidelli}, A., {Gomes}, R., {Backman}, D., 2007. {Planet
  Migration in Planetesimal Disks}. Protostars and Planets V, 669--684.

\bibitem[{{Lewis} and {Stewart}(2009)}]{LewisStewart2009}
{Lewis}, M.~C., {Stewart}, G.~R., Feb. 2009. {Features around embedded moonlets
  in Saturn's rings: The role of self-gravity and particle size distributions}.
  Icarus 199, 387--412.

\bibitem[{{Lin} and {Papaloizou}(1979)}]{LinPapaloizou1979}
{Lin}, D.~N.~C., {Papaloizou}, J., Mar. 1979. {Tidal torques on accretion discs
  in binary systems with extreme mass ratios}. \mnras 186, 799.

\bibitem[{{Masset}(2001)}]{Masset2001}
{Masset}, F.~S., Sep. 2001. {On the Co-orbital Corotation Torque in a Viscous
  Disk and Its Impact on Planetary Migration}. \apj 558, 453--462.

\bibitem[{{Masset} and {Papaloizou}(2003)}]{MassetPapaloizou2003}
{Masset}, F.~S., {Papaloizou}, J.~C.~B., May 2003. {Runaway Migration and the
  Formation of Hot Jupiters}. \apj 588, 494--508.

\bibitem[{{Paardekooper} et~al.(2009){Paardekooper}, {Baruteau}, {Crida}, and
  {Kley}}]{Paardekooper-etal-2009}
{Paardekooper}, S., {Baruteau}, C., {Crida}, A., {Kley}, W., Nov. 2009. {A
  torque formula for non-isothermal type I planetary migration - I. Unsaturated
  horseshoe drag}. \mnras, 1769--+.

\bibitem[{{Paardekooper} and
  {Papaloizou}(2009)}]{Paardekooper-Papaloizou-2009a}
{Paardekooper}, S., {Papaloizou}, J.~C.~B., Apr. 2009. {On corotation torques,
  horseshoe drag and the possibility of sustained stalled or outward
  protoplanetary migration}. \mnras 394, 2283.

\bibitem[{{Papaloizou} et~al.(2007){Papaloizou}, {Nelson}, {Kley}, {Masset},
  and {Artymowicz}}]{Papaloizou-etal-2007}
{Papaloizou}, J.~C.~B., {Nelson}, R.~P., {Kley}, W., {Masset}, F.~S.,
  {Artymowicz}, P., 2007. {Disk-Planet Interactions During Planet Formation}.
  In: Protostars and Planets V. pp. 655--668.

\bibitem[{{Porco} et~al.(2007){Porco}, {Thomas}, {Weiss}, and
  {Richardson}}]{Porco-etal-2007}
{Porco}, C.~C., {Thomas}, P.~C., {Weiss}, J.~W., {Richardson}, D.~C., Dec.
  2007. {Saturn's Small Inner Satellites: Clues to Their Origins}. Science 318,
  1602--.

\bibitem[{{Porco} et~al.(2004){Porco}, {West}, {Squyres}, {Alfred}, {Murray},
  {Del Genio}, {Ingersoll}, {Johnson}, {Neukum}, {Veverka}, {Dones}, {Brahic},
  {Burns}, {Haemmerle}, {Knowles}, {Dawson}, {Roatsch}, {Beurle}, and
  {Owen}}]{Porco-etal-2004}
{Porco}, C.~C., {West}, R.~A., {Squyres}, S.~M., {Alfred}, T.~P., {Murray},
  C.~D., {Del Genio}, A., {Ingersoll}, A.~P., {Johnson}, T.~V., {Neukum}, G.,
  {Veverka}, J., {Dones}, L., {Brahic}, A., {Burns}, J.~A., {Haemmerle}, V.,
  {Knowles}, B., {Dawson}, D., {Roatsch}, T., {Beurle}, K., {Owen}, W., Mar.
  2004. {Cassini Imaging Science: Instrument Characteristics And Anticipated
  Scientific Investigations At Saturn}. Space Science Reviews 115, 363--497.

\bibitem[{{Press} et~al.(1992){Press}, {Teukolsky}, {Vetterling}, and
  {Flannery}}]{NumRecipesF1992}
{Press}, W.~H., {Teukolsky}, S.~A., {Vetterling}, W.~T., {Flannery}, B.~P.,
  1992. {Numerical recipes in FORTRAN. The art of scientific computing}.
  Cambridge: University Press, |c1992, 2nd ed.

\bibitem[{{Rein} et~al.(2010){Rein}, {Lesur}, and
  {Leinhardt}}]{ReinLesurLeinhardt2010}
{Rein}, H., {Lesur}, G., {Leinhardt}, Z.~M., 2010. {The Validity of the
  Super-Particle Approximation during Planetesimal Formation}. \aap ,\ in
  press, ArXiv e--prints 1001.0109.

\bibitem[{Rein and Papaloizou(2009)}]{Rein-Papaloizou-2009}
Rein, H., Papaloizou, J.~C.~B., apr 2009. On the evolution of mean motion
  resonances through stochastic forcing: fast and slow libration modes and the
  origin of hd~128311. \aap 497~(2), 595--609.

\bibitem[{Rein and Papaloizou(2010)}]{Rein-Papaloizou-2010}
Rein, H., Papaloizou, J.~C.~B., 2010. Stochatic orbital migration of small
  bodies in saturn's rings. \aap, in preparation, available on arXiv.

\bibitem[{{Salo}(1995)}]{Salo1995}
{Salo}, H., Oct. 1995. {Simulations of dense planetary rings. III.
  Self-gravitating identical particles.} Icarus 117, 287--312.

\bibitem[{{Spahn} and {Srem{\v c}evi{\'c}}(2000)}]{SpahnSremcevic2000}
{Spahn}, F., {Srem{\v c}evi{\'c}}, M., Jun. 2000. {Density patterns induced by
  small moonlets in Saturn's rings?} \aap 358, 368--372.

\bibitem[{{Srem{\v c}evi{\'c}} et~al.(2007){Srem{\v c}evi{\'c}}, {Schmidt},
  {Salo}, {Sei{\ss}}, {Spahn}, and {Albers}}]{Sremcevic-etal-2007}
{Srem{\v c}evi{\'c}}, M., {Schmidt}, J., {Salo}, H., {Sei{\ss}}, M., {Spahn},
  F., {Albers}, N., Oct. 2007. {A belt of moonlets in Saturn's A ring}. \nat
  449, 1019--1021.

\bibitem[{{Tanaka} et~al.(2002){Tanaka}, {Takeuchi}, and
  {Ward}}]{Tanaka-etal-2002}
{Tanaka}, H., {Takeuchi}, T., {Ward}, W.~R., Feb. 2002. {Three-Dimensional
  Interaction between a Planet and an Isothermal Gaseous Disk. I. Corotation
  and Lindblad Torques and Planet Migration}. \apj 565, 1257--1274.

\bibitem[{{Tiscareno} et~al.(2008){Tiscareno}, {Burns}, {Hedman}, and
  {Porco}}]{Tiscareno-etal-2008}
{Tiscareno}, M.~S., {Burns}, J.~A., {Hedman}, M.~M., {Porco}, C.~C., Mar. 2008.
  {The Population of Propellers in Saturn's A Ring}. \aj 135, 1083--1091.

\bibitem[{{Tiscareno} et~al.(2006){Tiscareno}, {Burns}, {Hedman}, {Porco},
  {Weiss}, {Dones}, {Richardson}, and {Murray}}]{Tiscareno-etal-2006}
{Tiscareno}, M.~S., {Burns}, J.~A., {Hedman}, M.~M., {Porco}, C.~C., {Weiss},
  J.~W., {Dones}, L., {Richardson}, D.~C., {Murray}, C.~D., Mar. 2006.
  {100-metre-diameter moonlets in Saturn's A ring from observations of
  `propeller' structures}. \nat 440, 648--650.

\bibitem[{{Tiscareno et al.}(2010)}]{Tiscareno-etal-2010}
{Tiscareno et al.}, 2010. in preparation.

\bibitem[{{Toomre}(1964)}]{Toomre1964}
{Toomre}, A., May 1964. {On the gravitational stability of a disk of stars}.
  \apj 139, 1217--1238.

\bibitem[{{Ward}(1986)}]{Ward1986}
{Ward}, W.~R., Jul. 1986. {Density waves in the solar nebula - Differential
  Lindblad torque}. Icarus 67, 164--180.

\bibitem[{{Ward}(1991)}]{Ward1991}
{Ward}, W.~R., Mar. 1991. {Horsehoe Orbit Drag}. In: Lunar and Planetary
  Institute Conference Abstracts. pp. 1463--1464.

\bibitem[{{Ward}(1997)}]{Ward1997}
{Ward}, W.~R., Apr. 1997. {Protoplanet Migration by Nebula Tides}. Icarus 126,
  261--281.

\bibitem[{{Zebker} et~al.(1985){Zebker}, {Marouf}, and
  {Tyler}}]{Zebker-etal-1985}
{Zebker}, H.~A., {Marouf}, E.~A., {Tyler}, G.~L., Dec. 1985. {Saturn's rings -
  Particle size distributions for thin layer model}. Icarus 64, 531--548.

\end{thebibliography}

\end{document}